\documentclass[aip,reprint,onecolumn,amsmath,amssymb,groupedaddress,author-year]{revtex4-1}

\bibpunct{(}{)}{,}{a}{}{,}

\usepackage{amsmath}
\usepackage{amssymb}
\usepackage{subfigure}
\usepackage{graphicx}
\usepackage{color}
\usepackage{hyperref}
\usepackage{bm}

\DeclareMathOperator{\diag}{diag}		
\DeclareMathOperator{\trace}{tr}		

\begin{document}

\markboth{Desislava Chetalova et al.}{Portfolio return distributions}

\title{\Large Portfolio return distributions: Sample statistics with stochastic correlations}
\author{Desislava Chetalova}
\email{desislava.chetalova@uni-due.de}
\affiliation{Fakult\"at f\"ur Physik, Universit\"at Duisburg--Essen, Duisburg, Germany}
\author{Thilo A. Schmitt}
\affiliation{Fakult\"at f\"ur Physik, Universit\"at Duisburg--Essen, Duisburg, Germany}
\author{Rudi Sch\"afer}
\affiliation{Fakult\"at f\"ur Physik, Universit\"at Duisburg--Essen, Duisburg, Germany}
\author{Thomas Guhr}
\affiliation{Fakult\"at f\"ur Physik, Universit\"at Duisburg--Essen, Duisburg, Germany}
\date{\today}

\begin{abstract}
We consider random vectors drawn from a multivariate normal distribution and compute the sample statistics in the presence of stochastic correlations.
For this purpose, we construct an ensemble of random correlation matrices and average the normal distribution over this ensemble. The resulting 
distribution contains a modified Bessel function of the second kind whose behavior differs significantly from the multivariate normal distribution, in the central part as well as 
in the tails. 
This result is then applied to asset returns. We compare with empirical return distributions using daily data from the Nasdaq Composite Index in the period from 1992 to 2012. The comparison reveals good agreement, 
the average portfolio return distribution describes the data well especially in the central part of the distribution. This in turn 
confirms our ansatz to model the non-stationarity by an ensemble average.
\end{abstract}


\maketitle

\section{Introduction}

Financial markets are non-stationary. The non-stationarity manifests itself particularly in the fact that correlations change over time \citep[see e.g.,][]{Bekaert1995,Longin1995,Fenn2011,Muennix2012}. 
This has an impact on return distributions  of portfolios containing correlated financial instruments. We take the non-stationarity into account by a new approach based on random matrices. 
We emphasize that our random matrix approach is conceptually different from the previously observed random matrix behavior for correlation matrices. It was found \citep{Laloux1999, Plerou1999} that a large part of the eigenvalue density agrees with the Marchenko-Pastur law \citep{Marchenko1967}. 
The latter is due to ``noise dressing'', that is, it results from the finiteness of the time series. In contrast, our approach handles the problem of non-stationarity, as we discuss in detail.

In case of correlated returns one is interested in their joint distribution which contains the whole information about the individual distributions as well as their dependence structure. Here, we derive a multivariate distribution to model jointly the stock returns of a correlated market considering the fluctuation of the correlations. 
More precisely, we address the question what sample statistics to expect if we assume a multivariate normal distribution for each realization of a sample, but with a randomly drawn covariance or correlation matrix. The assumption of joint normality is justified on short time scales where the covariance matrix can be viewed as fixed \citep{Schmitt2013}.
We derive the sample statistics analytically by averaging over an ensemble of Wishart random correlation matrices. The resulting multivariate probability density function (pdf)
can be expressed in terms of a Macdonald function, a modified Bessel function of the second kind. It depends on the average covariance matrix and a single parameter controlling the variance of the random matrix ensemble. 
The new multivariate distribution belongs to the broad family of elliptical distributions \citep{Cambanis1981} which generalize the multivariate normal distribution while inheriting many of its useful properties.
Moreover, it includes several heavy-tailed distributions, which makes it very attractive for modeling of financial data. 
In particular, the multivariate Student's $t$-distribution has received much attention in the context of modeling multivariate financial returns \citep[see e.g.,][]{Breymann2003, Mashal2003}.

Our approach is related to but different from a compounding or mixture approach of univariate distributions \citep[see][]{dubey1970compound,Clark1973,Barndorff-Nielsen1982}. These approaches are motivated by the empirical observation of volatility fluctuations \citep[see e.g.,][]{black76,christie82,tauchen83} and by autoregressive models such as GARCH \citep{Engle1982,Bollerslev1986} and EGARCH \citep{Nelson1991}, which aim to capture these fluctuating volatilities.

We point out that we propose a model for the unconditional return distributions, taking into account the non-stationarity of the correlations by an ensemble of random matrices. This is different from the multivariate GARCH models \citep[see e.g.,][]{ Engl2002, Tse2002, Golosnoy2012}
where correlations are modeled by stochastic processes.

The paper is organized as follows. In Sec.~\ref{section2} we present the random matrix approach and the analytical computation of the sample statistics for the general case of a random sample with multivariate normally distributed realizations. In Sec.~\ref{section3} we transfer our findings to financial data and apply our results to the sample statistics of asset returns. 
Here we assume normally distributed stock returns with stochastic correlations. 
We derive a univariate pdf for a portfolio return and compare it with empirical data for the Nasdaq Composite stocks in the observation period $1992-2012$ in Sec.~\ref{section4}. We conclude our findings in Sec.~\ref{section5}.

\section{Correlation Averaged Normal Distribution} \label{section2}

Consider a sample of $K$ dimensional random vectors $x$ each drawn from a multivariate normal distribution with the pdf
\begin{align}
g ( x| \Sigma_{\rm s}) & = \frac{ 1 }{ \sqrt{2 \pi }^K } \frac{ 1 }{ \sqrt{\det \Sigma_{\rm s} } } \exp \left( - \frac{ 1 }{ 2 }  x^\dagger \Sigma^{-1}_{\rm s} x \right)  \ ,
\label{eq:multidist}
\end{align}
where $\Sigma_{\rm s}$ is the covariance matrix of the realization $x$ and $^\dagger$ denotes the transpose. For later purposes, we write this pdf as a Fourier transform
\begin{align}
g (x| \Sigma_{\rm s}) = \frac{ 1 }{ (2 \pi)^K } \int {d} [\omega] \ { e}^{ - { i} \omega^\dagger  x } \exp \left( - \frac{1}{2}  \omega^\dagger \Sigma_{\rm s} \, \omega \right)  \ ,
\label{eq:fouriertr}
\end{align}
where $\omega $ is a $K$ component real vector and the measure $ {d}[\omega] $ is the product of the differentials of the individual elements. 
The integration domain is always the entire real axis.

The covariance matrix changes between observations. We take this into account by replacing the covariance matrix with a random matrix
\begin{align}
\Sigma_{\rm s} \quad \longrightarrow \quad \sigma W W^\dagger \sigma \ ,
\label{eq:rcm}
\end{align}
where the $ K \times K $ diagonal matrix $\sigma$ contains the standard deviations ${\sigma}=\diag(\sigma_1, \dots, \sigma_K)$ for each random variable. 
The elements of the $ K\times N $ rectangular matrix $ W $ are drawn from a Gaussian distribution with the pdf
\begin{align}
w ({W}| {C},N) = \sqrt { \frac{ N }{ 2 \pi } }^{ K N } \frac{ 1 }{ \sqrt{\det { C} }^N } \exp \left( - \frac{ N }{ 2 } \trace {W}^\dagger {C}^{-1} {W} \right) \ , 
\label{eq:gauss}
\end{align}
where $C$ is the average correlation matrix. Thus, we construct an ensemble of random correlation matrices $ WW^\dagger $ which follow a Wishart distribution of the form \citep{Wishart1928}
\begin{align}
\tilde w ( {X}| {C},N ) =\frac{ \sqrt{N}^{K N} \sqrt{\det {X} }^{ N - K - 1} }{ \sqrt{ 2 }^{ K N }  \Gamma_K (N/2) \sqrt{ \det C }^N }   \exp \left( - \frac{N}{2} \trace C^{-1} { X} \right)  
\label{eq:wish}
\end{align}
with $X\equiv WW^\dagger$ and the multivariate Gamma function  $\Gamma_K (\cdot)$. The Wishart correlation matrix ensemble fluctuates around the average correlation matrix $ C$.
By construction, the ensemble average of the model correlation matrix equals $C$,
\begin{align}
\left< {WW}^\dagger \right> = \int {d}[ {W} ] \ w ({ W}| {C},N ) \ {W W}^\dagger = {C} \ ,
\end{align}
where the measure $ {d}[W] $ is the product of the differentials of the matrix elements.
We point out that the parameter $N$ is proportional to the inverse variance of the distribution~(\ref{eq:wish})
\begin{align}
 {\rm var} ({X}_{ij})=\frac{c_{ij}^2+c_{ii} c_{jj}}{N} \ ,
\end{align}
where $c_{ij}$ is the $ij$-th element of the average correlation matrix $C$. 
The larger $N$, the narrower the distribution of the elements of $W W^\dagger$ becomes. In the limit $N\rightarrow \infty$ the random correlation matrix $WW^\dagger$ is fixed without fluctuations.
We obtain an invertible random correlation matrix in the case $N \geq K$. For $N < K$, however, the resulting matrix is not invertible. Nevertheless, the pdf~(\ref{eq:multidist}) is well defined in terms of proper $\delta$ functions, as discussed in \ref{appA}.
 
Instead of the correlation matrix we can express the full covariance matrix $\Sigma=\sigma  C \sigma$ as a random matrix
\begin{align}
\Sigma_{\rm s} \quad \longrightarrow \quad A A^\dagger  
\end{align}
with $ A A^\dagger$ being a Wishart random matrix. This leads to the same result, as discussed in \ref{appB} and explicitly shown in \citet{Schmitt2013}. Mathematically, it does not make a difference whether we perform the calculation with a 
random covariance or a random correlation matrix. Thus, our approach does not contradict the empirical observation of fluctuating volatilities.

Our key idea is to average the multivariate normal distribution~(\ref{eq:multidist}) with the random covariance matrix~(\ref{eq:rcm}) over the Gaussian distribution~(\ref{eq:gauss}),
\begin{align}
\langle  g \rangle ({x}| {C},N) = \int  {d}[ {W} ] \ w ({W}| {C},N) \ g (x| {\sigma W W^\dagger \sigma}) \ .
\label{eq:ansatz}
\end{align}
Inserting the Fourier transform~(\ref{eq:fouriertr}) into Eq.~(\ref{eq:ansatz}) leads to
\begin{align}
\langle  g \rangle ({ x}| {C},N) &= \int {d}[ {W} ] \ \sqrt{ \frac{ N }{ 2 \pi} }^{ K N } \exp \left( - \frac{ N }{ 2 } \trace {W}^\dagger {C}^{-1} {W} \right) \notag \\ 
& \times \frac{ \sqrt { \det { C} }^{ -N } }{ ( 2 \pi )^K } \int {d}[\omega] \ {e}^{ - { i}  \omega^\dagger  x } \exp \left( - \frac{ 1 }{ 2 } \omega^\dagger \sigma {W W}^\dagger \sigma  \omega \right) .
\end{align}
We rewrite the scalar bilinear form in the exponent as  $\omega^\dagger \sigma { W W}^\dagger \sigma \omega =  \trace ( { W}^\dagger \sigma \omega  \omega^\dagger \sigma {W})$. We merge the two traces and arrive at
\begin{align}
\langle  g \rangle ({x}| {C},N) &= \sqrt{ \frac{ N }{ 2 \pi} }^{ K N } \frac{ \sqrt { \det {C} }^{ -N } }{ ( 2 \pi )^K } \int {d}[\omega] \ {e}^{ -{i}  \omega^\dagger x } \notag \\
& \times \int {d}[ {W} ] \ \exp \left( - \frac{ 1 }{ 2 } \trace \left( {W W}^\dagger \left( N {C}^{-1} +   \sigma \omega \omega^\dagger \sigma \right) \right) \right) \ .
\end{align}
Here, and in similar cases later on, we may exchange the order of integration as the Fourier representation~(\ref{eq:fouriertr}) is robust in a distributional sense while the Gaussian distribution does not inflict any convergence problems. Since the integral over $ W $ is simply Gaussian we have
\begin{align}
\langle  g \rangle ({x}| {C},N) & = \frac{ \sqrt{ N }^{ K N } \sqrt { \det {C} }^{-N} }{ ( 2 \pi )^K } \int {d}[\omega] \ {e}^{ -{i}  \omega^\dagger x} \frac{ 1 }{ \sqrt{\det (N {C}^{-1} + \sigma  \omega \omega^\dagger \sigma)}^N } .
\end{align}
As $\sigma \omega \omega^\dagger \sigma$ is a dyadic matrix of rank unity we find
\begin{align}
\det (N {C}^{-1} + \sigma \omega \omega^\dagger \sigma ) =  N^K (1 +  \omega^\dagger \sigma {C} \sigma  \omega / N)  \det { C}^{-1}   ,
\end{align}
which implies
\begin{align}
\langle  g \rangle ({x}| {C},N) = \frac { 1 }{ (2 \pi)^K } \int {d}[ \omega ] \ { e}^{ -{ i}  \omega^\dagger  x } \frac { 1 }{ \sqrt{ 1 +  \omega^\dagger \sigma C \sigma \omega / N }^N }  \ . 
\label{eq:eq6}
\end{align}
We use the formula
\begin{align}
\frac{ 1 }{ a^\eta } = \frac{ 1 }{ \Gamma(\eta)} \int\limits^\infty_0 z^{\eta - 1} {e}^{-a z} \, { d} z
\end{align}
for real and positive variables $a$ and $\eta$. We identify $a$ with the radicand of the square root and $\eta$ with $N/2$, and cast Eq.~(\ref{eq:eq6}) into the form
\begin{align}
\langle  g \rangle ({x}| { C},N) &= \frac{1}{ (2 \pi)^K \Gamma( N/2 ) } \int\limits_0^\infty {d}z \ z^{\frac{ N }{ 2 } - 1 } {e}^{-z} \int {d}[\omega] \ {e}^{ -{ i}  \omega^\dagger x } \exp \left( - \frac{ z }{ N } \omega^\dagger \sigma {C} \sigma \omega \right) .
\end{align}
By applying the inverse Fourier transform we end up with
\begin{align}
\langle  g \rangle ({ x}| \Sigma, N) &= \frac{1}{ (2 \pi)^K \Gamma( N/2 ) \sqrt{ \det \Sigma } } \int\limits_0^\infty {d} z \ z^{\frac{ N }{ 2 } - 1 } {e}^{-z} \sqrt{ \frac{ \pi N }{ z } }^K \exp \left( - \frac{ N }{ 4 z }  x^\dagger \Sigma^{-1} x \right) .
\label{eq:result}
\end{align}
Here we introduce the new, fixed matrix 
\begin{align}
\Sigma= \sigma C \sigma \quad ,
\end{align}
which represents the average covariance matrix in the sense that $ C $ is the average correlation matrix and $ \sigma $ the diagonal matrix of the standard deviations. We note that Eq.~(\ref{eq:result}) can be expressed as an average of the 
Gaussian with an argument $x \Sigma^{-1} x $ over the variance $z$ which is $\chi^2$ distributed, see \citet{Schmitt2013}  and \ref{appC} for more details. Finally, we perform the integral over $z$ and obtain the pdf
\begin{align}
\langle  g \rangle ({x}| \Sigma, N) & = \sqrt{ \frac{ N }{ 4 \pi } }^K \frac{ \sqrt{2}^{ K-N+2 } }{ \Gamma(N/2) } \frac{ \sqrt{ N  x^\dagger \Sigma^{-1}  x}^{\frac{N - K }{ 2 } } }{ \sqrt{ \det \Sigma} } \mathcal{K}_{\frac{K-N}{2}} \left( \sqrt{ N  x^\dagger \Sigma^{-1}  x} \right) ,
\label{eq:mainresult}
\end{align}
where $\mathcal{K}_\nu(\cdot)$ is the modified Bessel function of the second kind of order $\nu$.
The pdf differs significantly from the multivariate normal distribution. It depends only on the average covariance matrix $\Sigma$ and 
the free parameter $N$ which characterizes the fluctuations around $\Sigma$. 
Furthermore, due to the invariance of the Wishart distribution, the vector $ x$ enters the result only via the bilinear form $  x^\dagger \Sigma ^{-1} x $. 
The tails decay exponentially. The smaller $N$, the heavier the tails become. In the limit $N \rightarrow \infty$ the pdf converges towards the multivariate 
normal distribution. The marginal distributions are $\mathcal K$ Bessel functions as well, of the order $(N-1)/2$, see \citet{Schmitt2013}.

We note that our model is reminiscent of the Bayesian analysis. In particular, \citet{Bekker1995} consider the Bayesian analysis of the multivariate normal distribution using a Wishart covariance prior and provide calculations with results which are formally similar to ours. However, the interpretation is completely different since in our model each realization is drawn from a different distribution.
Moreover, we go one step further by reducing the result to a Bessel function with scalar argument.

\section{Application to Asset Returns} \label{section3}

We now apply our model to financial return time series. Consider a portfolio consisting of $K$ assets. The return time series are given by
\begin{align}
 r_k (t) = \frac{ S_k(t + \Delta t)-S_k(t) }{ S_k(t) } \quad ,
\end{align}
where $ S_k(t) $ is the price of the $ k $-th asset at time $ t $ and $ \Delta t $ is the return interval. 

As already mentioned, we use the common assumption of normally distributed asset returns in order to apply our model to financial data.
Under this assumption the pdf of the return vector 
\begin{align*} 
r (t) = \left( r_1 (t),\dots ,r_K (t) \right)
\end{align*}
at each fixed time $ t $ can be described by Eq.~(\ref{eq:multidist}) but with different covariance matrices since correlations between the individual assets change in time. 
However, if we are interested in the pdf of the return vector in the whole observation period, we have to average over all correlation matrices. By applying the result from the previous section we obtain the average return pdf
\begin{align}
\langle  g \rangle ({ r}| \Sigma, N)&= \frac{ 1 }{ (2 \pi)^K \ \Gamma(N/2) \sqrt{ \det \Sigma } } \int\limits_0^\infty {d}z \ z^{ \frac{ N }{ 2 } - 1 } { e}^{-z} \sqrt{ \frac{ \pi N }{ z } }^K \exp \left( -\frac{ N }{ 4 z }  r^\dagger \Sigma^{-1} r \right) \quad ,
\label{eq:avretdist}
\end{align}
where $ \Sigma =  \sigma C \sigma $ is the average empirical covariance matrix calculated over the whole observation period.
To simplify the notation we drop the argument $ t $ in the returns, but keep in mind that they are always measured at one given time $ t $ and depend on the return interval $ \Delta t $.

In the following we will concentrate on the implications of this result for portfolio returns. A comparison of the multivariate distribution~(\ref{eq:avretdist}) with data has been carried out elsewhere \citep{Schmitt2013}, with a discussion aimed at a physics audience.
The comparison reveals a good agreement between model and data, in particular in the central part of the distribution. There are, however, small deviations in the tails. 

We now employ the result~(\ref{eq:avretdist}) to compute the average pdf of a portfolio return. 
The return of a portfolio is given by the weighted sum of the individual asset returns
\begin{align}
R = \sum_{k=1}^K u_k r_k =   u^\dagger  r \quad ,
\end{align}
where $ u_k $ is the weight of the $ k $-th asset. The weights obey the normalization condition $ \sum_{k=1}^K u_k =1 $. 

\begin{figure}
  \centering
   \includegraphics[width=\textwidth]{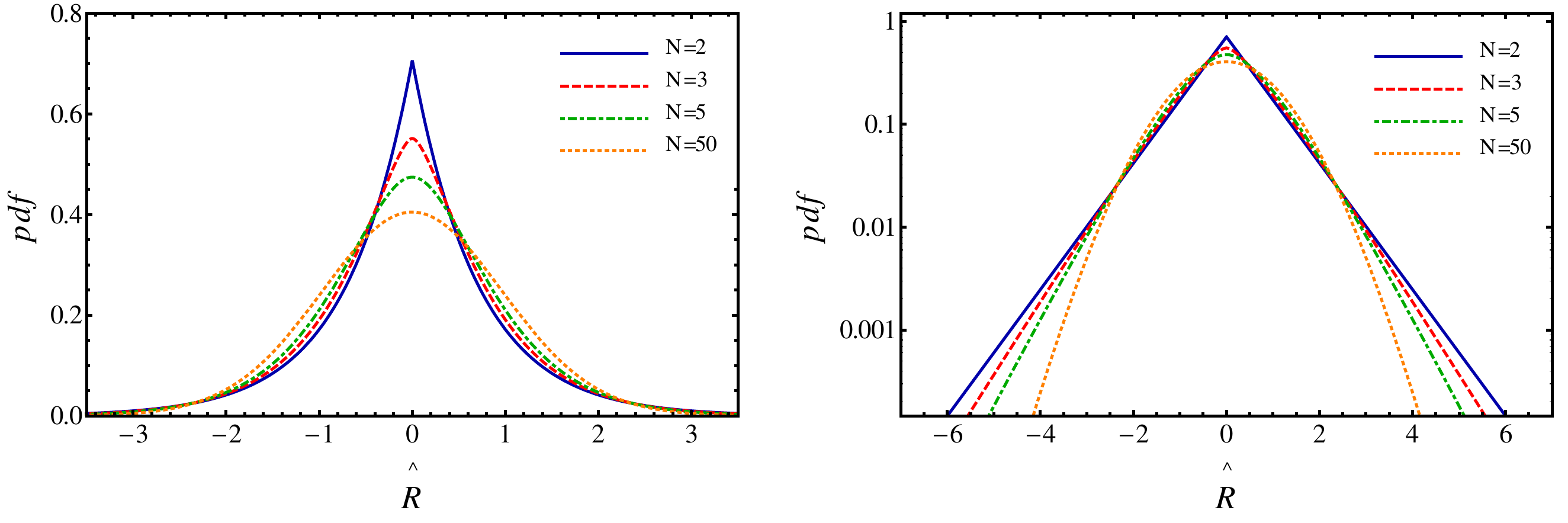} 
  \caption{Illustration of the rescaled average portfolio return pdf $\langle f \rangle (\widehat R| N)$ for different values of $N$, plotted linearly (left) and logarithmically (right). Solid, dashed, dashed-dotted and 
dotted lines correspond to $N=2, 3, 5, 50$, respectively.}
\label{fig1}
\end{figure}

The average portfolio return pdf is then
\begin{align}
\langle f \rangle (R | \Sigma, N) & = \int {d} [ r ] \langle  g \rangle ({r}| \Sigma, N) \ \delta \left( R -  u^\dagger  r \right) \\
& = \frac{ 1 }{ 2 \pi } \int\limits_{ -\infty}^{ +\infty} {d}\nu \ {e}^{ -{i} \nu R } \int d [ r ] \langle  g \rangle ({r}| \Sigma, N)  \ {e}^{ {i} \nu  u^\dagger  r } \ ,
\end{align}
with the integral over $ r $ being the characteristic function.
Inserting the pdf~(\ref{eq:avretdist}) we find
\begin{align}
\langle f \rangle (R | \Sigma, N) & = \frac{ 1 }{ (2 \pi)^{K+1} \ \Gamma(N/2) \sqrt{ \det \Sigma } } \int\limits_0^\infty { d}z \ z^{ \frac{ N }{ 2 }-1 } { e}^{-z} \sqrt{ \frac{ N \pi }{ z } }^K \notag \\
& \times  \int\limits_{ -\infty}^{ +\infty} {d} \nu \ {e}^{ -{i} \nu R } \int {d}[ r ] \exp \left( { i} \nu  u^\dagger  r -\frac{ N }{ 4 z } r^\dagger \Sigma^{-1} r \right) .
\end{align}
The integral over $ r $ is Gaussian. Carrying out the integration we find that the integral over $\nu$ is Gaussian as well and we have 
\begin{align}
\langle f \rangle (R | \Sigma, N) = \frac{ 1 }{ \Gamma(N/2) } \sqrt{ \frac{ N }{ 4 \pi  u^\dagger \Sigma  u} } \int\limits_0^\infty {d}z \ z^{ \frac{ N-3 }{ 2 } }  { e}^{-z} \exp \left( -\frac{ N R^2 }{ 4 z  u^\dagger \Sigma  u} \right) \ .
\end{align}
Once more, the last integral is a representation of the modified Bessel function of the second kind, this time of the order $ (N-1)/2 $. Thus, we obtain
\begin{align}
\langle f \rangle (R | \Sigma, N) &= \frac{ \sqrt{ 2 }^{ 1-N } }{ \sqrt{ \pi } \ \Gamma(N/2) }  \sqrt{ \frac{ N }{  u^\dagger \Sigma u} }^{ \frac{ N+1 }{ 2 } } \left|R \right|^{ \frac{ N-1 }{ 2 } } \ \mathcal K_{ \frac{ N-1 }{ 2 } }\left( \left| R\right| \sqrt{ \frac{ N }{  u^\dagger \Sigma  u } } \right) 
\label{eq:avprd}
\end{align}
for the average pdf of a portfolio return. It depends only on the free parameter $ N $, which describes the fluctuation strength of the correlations in the observation period, 
and on the scale variable 
\begin{align}
\alpha =  u^\dagger \Sigma u \quad ,
\end{align}
which can be computed from the portfolio weights and the fixed empirical covariance matrix $\Sigma$. 
We notice that $ \alpha $ is the bilinear form of the weights $  u $ with the covariance matrix $ \Sigma $ and thus has a very direct economic relevance as one single effective variance parameter of the entire portfolio. 
Thus, if we normalize the portfolio return $ R $ in the following way
\begin{align}
\widehat R = \frac{ R }{ \sqrt{ \alpha } } \quad ,
\label{eq:rescaledportfreturn}
\end{align}
we end up with the pdf 
\begin{align}
\langle f \rangle (\widehat R | N) &= \frac{ \sqrt{ 2 }^{ 1-N } }{ \sqrt{ \pi } \ \Gamma(N/2) } \sqrt{ N } ^{ \frac{ N+1 }{ 2 } } |\widehat R |^{ \frac{ N-1 }{ 2 } } \ \mathcal K_{ \frac{ N-1 }{ 2 } }\left( | \widehat R | \sqrt{ N } \right) ,
\label{eq:avprd2}
\end{align}
in which $N$ is the only free parameter. We illustrate this density function in Fig.~\ref{fig1} for different values of $N$.
Note that the tails become heavier as the value of $N$ decreases.
For large values of $N$ the pdf approaches the normal distribution.

\section{Comparison with Empirical Portfolio Returns} \label{section4}

\begin{figure}
  \centering
   \includegraphics[width=\textwidth]{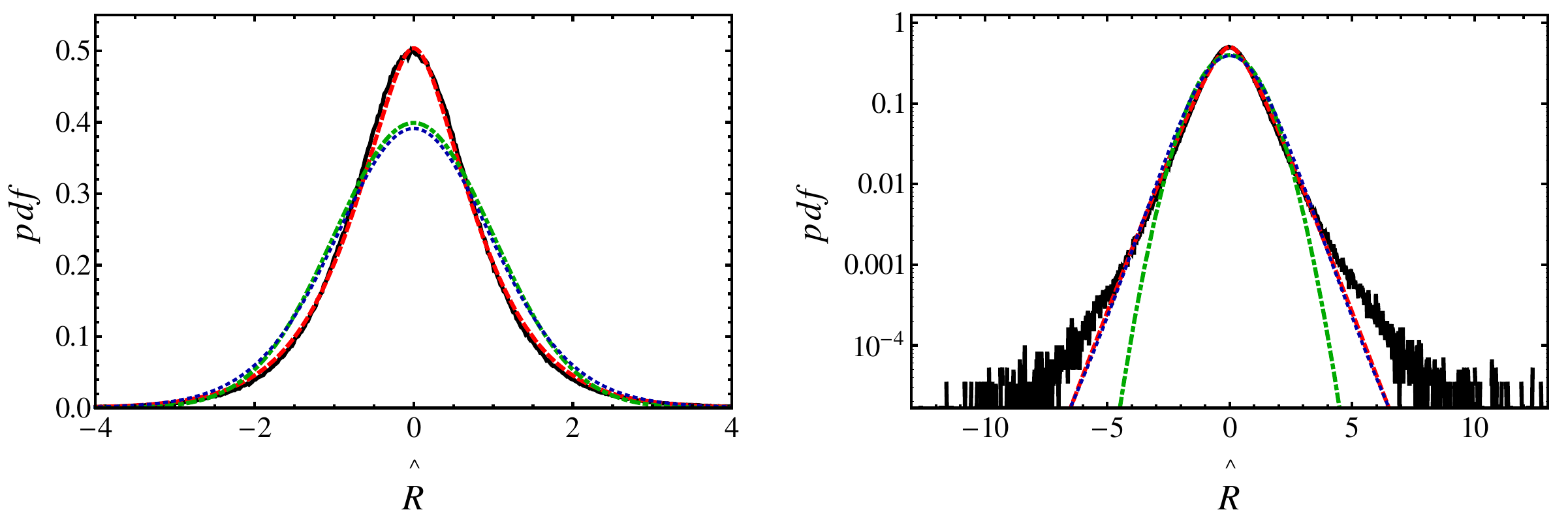} 
  \caption{Histogram of the rescaled empirical portfolio return $\widehat R$ (solid black) for daily returns and weights $u_k \sim \mathcal U (-0.5,0.5)$
compared with the average portfolio return pdf $\langle f \rangle (\widehat R | N) $ (dashed red) with $N=3.9$, plotted linearly (left) and logarithmically (right). The green dashed-dotted line shows a normal distribution $\mathcal N (0,1)$ and the blue dotted line a
Student's $t$-distribution $t(12.73)$.} 
\label{fig2}
\end{figure}

To compare our model with empirical data we consider the Nasdaq Composite Index in the period from January 1992 to March 2012.
We take into account $241$ stocks, for which the time series are complete and cover the whole observation period. 
To compute the empirical portfolio return we build portfolios each consisting of $ K $ stocks, which are randomly chosen from all available stocks. 
For each stock we calculate the stock returns on a given return interval, for each portfolio the corresponding portfolio return. 
Fig.~\ref{fig2} shows the histogram of the rescaled empirical portfolio return $\widehat R $ taking into account $600$ random portfolios of $20$ stocks each. 
We use daily returns, $\Delta t = 1 \ \mbox{day}$, and consider positive and negative weights drawn from a symmetric uniform distribution $ \mathcal U (-a,a) $ with $a=0.5$ in such a way that the normalization condition is satisfied.
Compared with the normal distribution $(\mu=0, \ \sigma=1)$, the histogram has a higher peak around zero and fatter tails. The Student's $t$-distribution ($\nu=12.73$) describes the tails much better than the normal distribution, however it 
fails to describe the center of the histogram. The parameters of the normal and the Student's $t$-distributions are estimated with the maximum likelihood method.
Compared with both distributions the average portfolio return pdf~(\ref{eq:avprd2}) resembles the data much better in the center of the histogram. In the tails it matches the Student's $t$-distribution. 
We point out that although our analytical result was derived for integer values of $N$, we can easily extend this result to real values. Allowing real values for $N$, we obtain the best agreement for $N=3.9$. 
Still, there are some deviations in the tails.
We determine the free parameter $ N $ with a minimum distance estimation method using a weighted Cramer-von Mises statistics which is often applied to accentuate the distance between model and empirical distribution in those parts of the distribution where more sensitivity is desired \citep{Parr1981, Boos1982, Ozturk1997}. Here, we aim at fitting particularly the  
center of the distribution since there the best match between model and data is observed. To give more weight to the center of the distribution, we use a Gaussian function of the form $\exp \left( -y^2/2 c^2 \right)$ with $c= 0.07$ as a weighting function. The maximum likelihood method yields about $10 \%$ smaller values for the parameter $N$ since it also takes the tails into account.

\begin{figure}
  \centering
   \includegraphics[width=\textwidth]{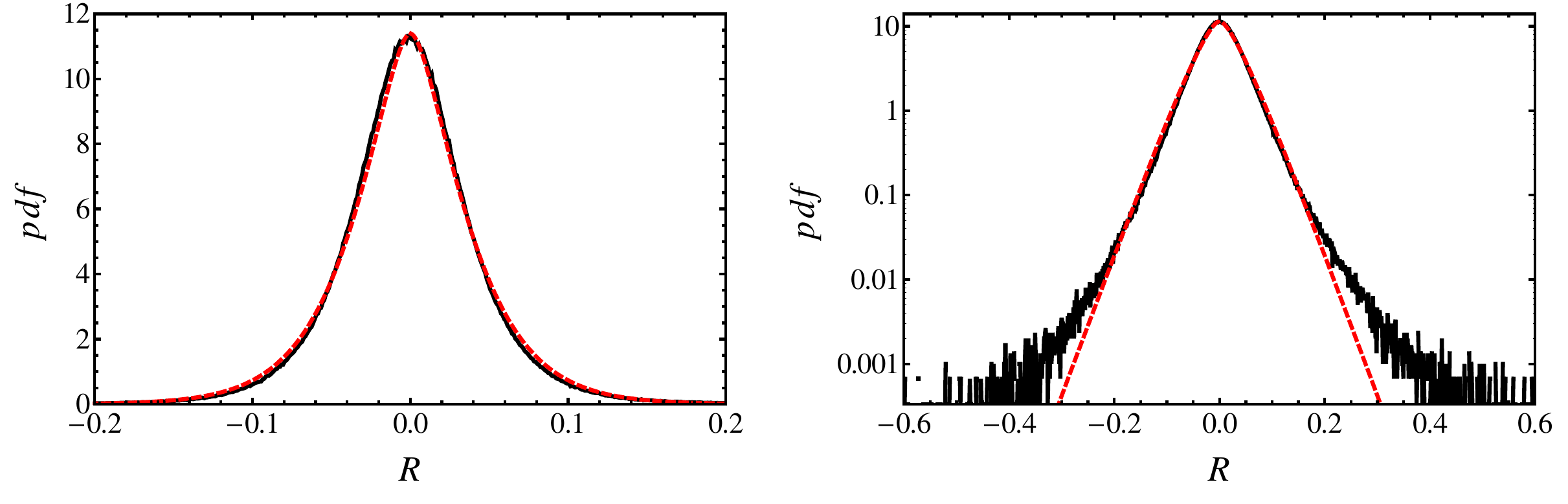} 
  \caption{Histogram of the  empirical portfolio return $R$ (solid black) for daily returns and weights $u_k \sim \mathcal U (-0.5,0.5)$
compared with the average portfolio return pdf $\langle f \rangle (R | \Sigma, N) $ (dashed red) with $\alpha=2.09 \times 10^{-3}$ and $N=3.5$, plotted linearly (left) and logarithmically (right).}
\label{fig3}
\end{figure}

\begin{figure}
  \centering
   \includegraphics[width=\textwidth]{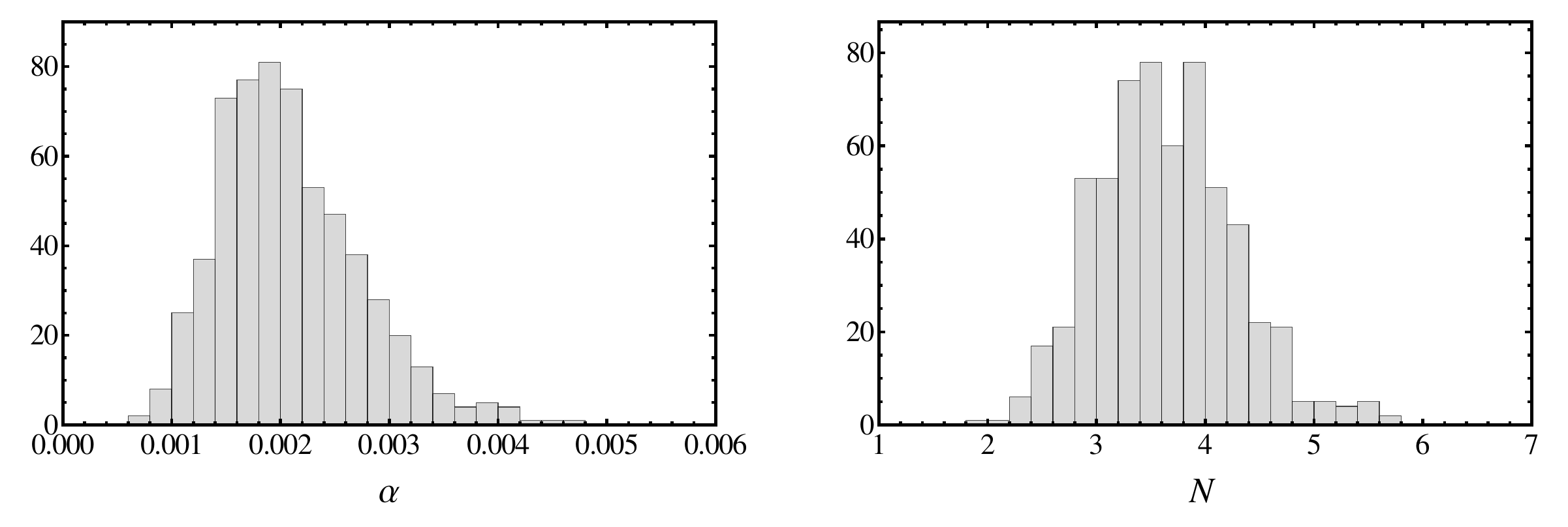} 
  \caption{Histograms of the portfolio variance $\alpha$ (left) and of the $N$ value (right) for each portfolio for the case $u_k \sim \mathcal U (-0.5,0.5)$.} 
\label{fig4}
\end{figure}

\begin{figure}
  \centering
   \includegraphics[width=\textwidth]{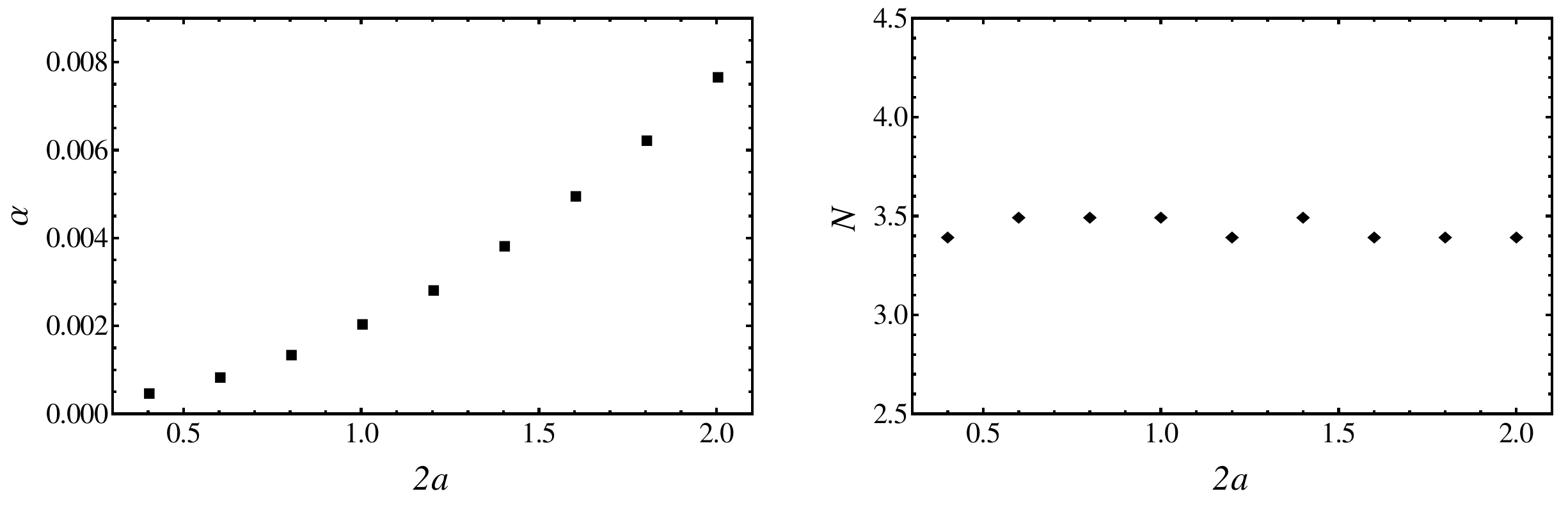}  
  \caption{ The parameters $\alpha$ (left) and $N$ (right) in dependence of the uniform distribution width $2a$.  }
\label{fig5}
\end{figure}

In the following we will consider the portfolio return $R$ directly, as it is the economically relevant quantity, and will investigate the impact of the portfolio weights. Fig.~\ref{fig3} shows the histogram
of the empirical portfolio return $ R $ compared with the average portfolio return pdf~(\ref{eq:avprd}). 
Again we consider daily returns and  weights drawn from a symmetric uniform distribution $ \mathcal U (-a,a) $ with $a=0.5$.  The minimum distance estimation method yields $N=3.5$.
The portfolio variance $ \alpha $ is computed for each portfolio from the corresponding weights and covariance matrix and then averaged over all available portfolios. 
The theoretical curve agrees well with the histogram in the central part, there are some deviations in the tails. Fig.~{\ref{fig4}} shows the range of $\alpha$ and $N$ values induced by the different portfolios. 
Choosing different distribution parameters for the portfolio weights affects the portfolio variance $\alpha$, see Fig.~{\ref{fig5}}. It increases monotonically with the distribution width $2a$. The parameter $N$, on the other hand, remains nearly constant. 

\begin{figure}
  \centering
   \includegraphics[width=\textwidth]{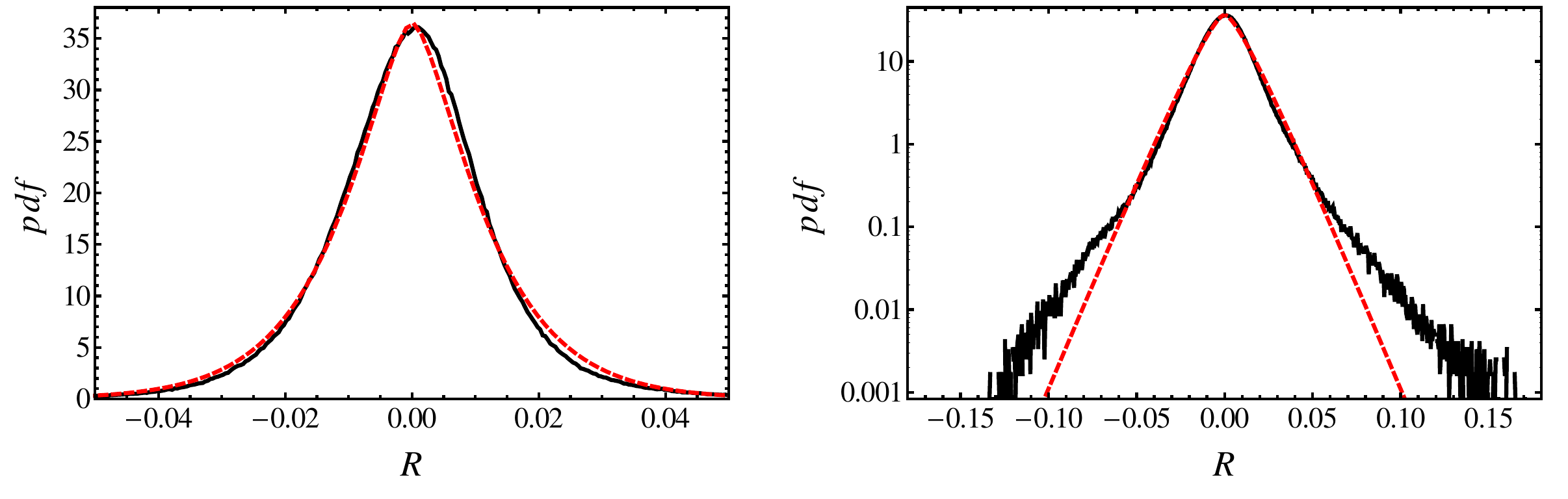}  
  \caption{Histogram of the  empirical portfolio return $ R $ (solid black) for daily returns and equal weights $u_k=1/K$ compared with the average portfolio return
pdf $\langle f \rangle (R | \Sigma, N) $ (dashed red) with $\alpha=2.17 \times 10^{-4}$ and $N=3.2$, plotted linearly (left) and logarithmically (right).}
\label{fig6}
\end{figure}

We now examine the case when all weights are positive, i.e., when short selling is not allowed. Fig.~\ref{fig6} shows the histogram of the empirical portfolio return $ R $ using daily returns and equal weights $ u_k=1/K $.
The histogram is asymmetric with a heavier tail on the right hand side.
Although the deviations in the tails are slightly more pronounced, the average return pdf with $N=3.2$ still agrees well with the histogram in the central part.
The portfolio variance, on the other hand, is much smaller than in the case of uniform portfolio weights, which implies a lower risk for positively weighted portfolios.

\begin{figure}
  \centering
   \includegraphics[width=\textwidth]{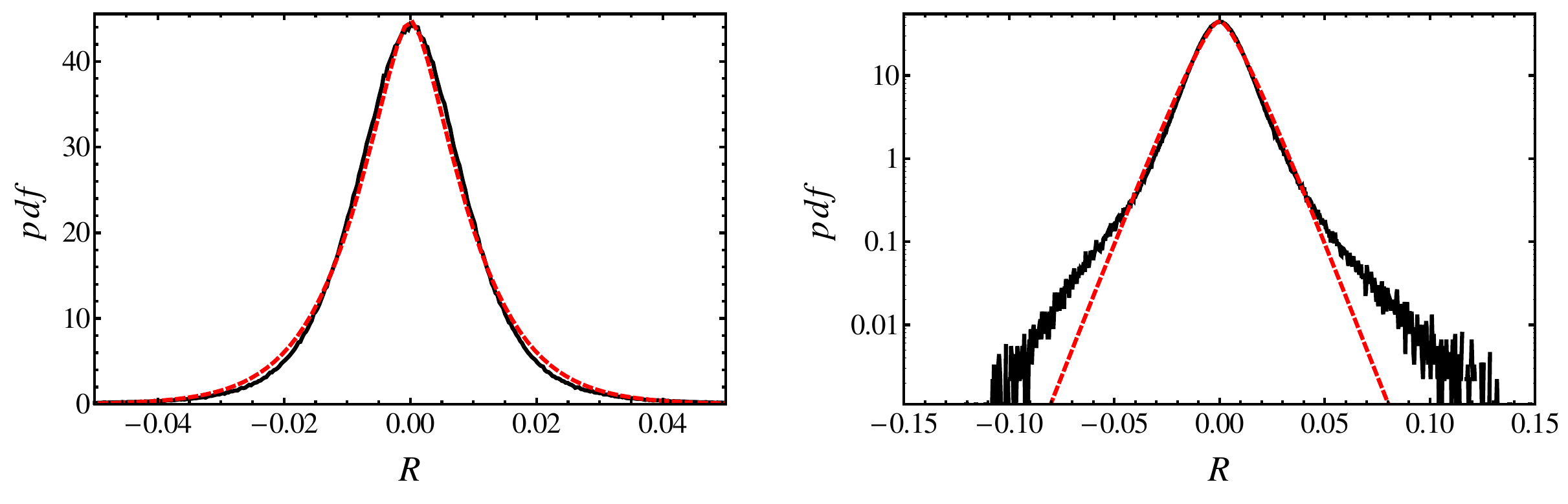}  
\caption{Histogram of the  empirical portfolio return $R$ (solid black) for daily returns and optimal weights
compared with the average portfolio return pdf $\langle f \rangle (R | \Sigma, N) $ (dashed red) with $\alpha=1.38 \times 10^{-4}$ and $N=3.4$, plotted linearly (left) and logarithmically (right).}
\label{fig7}
\end{figure}

Instead of drawing random weights, we can determine a set of optimal portfolio weights according to \citet{Markowitz1952}
\begin{align}
 u_{{\rm opt}}=\frac{\Sigma^{-1}  g}{ g^{\dagger} \Sigma \  g}
\label{eq:opt_weights}
\end{align}
with the $K$ dimensional vector $g=(1,\dots, 1)$ and the covariance matrix  $\Sigma$, computed for a given subset of stocks in the whole observation period. Using the optimal portfolio weights, we obtain the minimal portfolio variance. The histogram of the empirical portfolio return $R$, compared with the 
average portfolio return pdf with $N=3.4$, is shown in Fig.~\ref{fig7}. The histogram is asymmetric with a heavier tail on the right hand side. Again we observe a good agreement in the central part and deviations in the tails. Indeed, we find the smallest portfolio variance $\alpha$, about a factor $1.5$ smaller than the second best $\alpha$ for $u_k=1/K$.

\begin{figure}
  \centering
   \includegraphics[width=\textwidth]{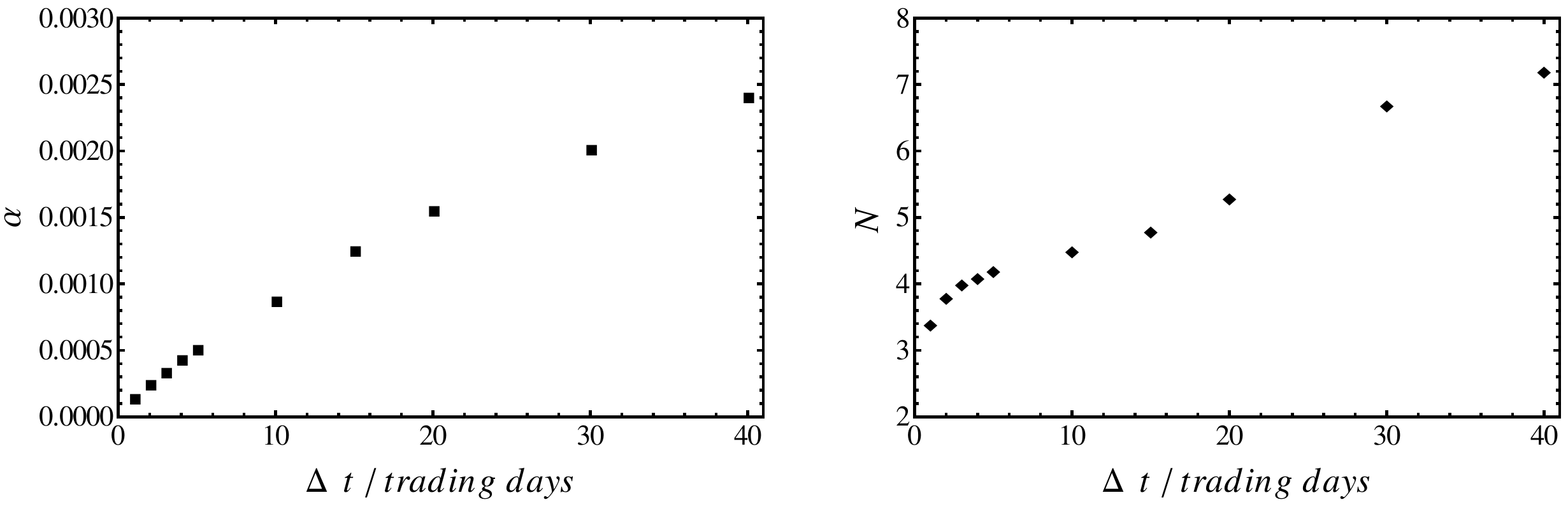}  
  \caption{The parameters $\alpha$ (left) and $N$ (right) in dependence of the return interval $\Delta t$ using optimal weights.}
\label{fig8}
\end{figure}

So far, we have only considered daily returns, $\Delta t = 1 \ \mbox{day}$. Let us now take a look at other return intervals. 
Fig.~\ref{fig8} shows the parameters $\alpha$ and $N$ for different return intervals between one day and two months. We observe that $N$ increases with $\Delta t$, which leads to a more Gaussian-like distribution. The portfolio variance increases too. 

\begin{figure}
  \centering
   \includegraphics[width=\textwidth]{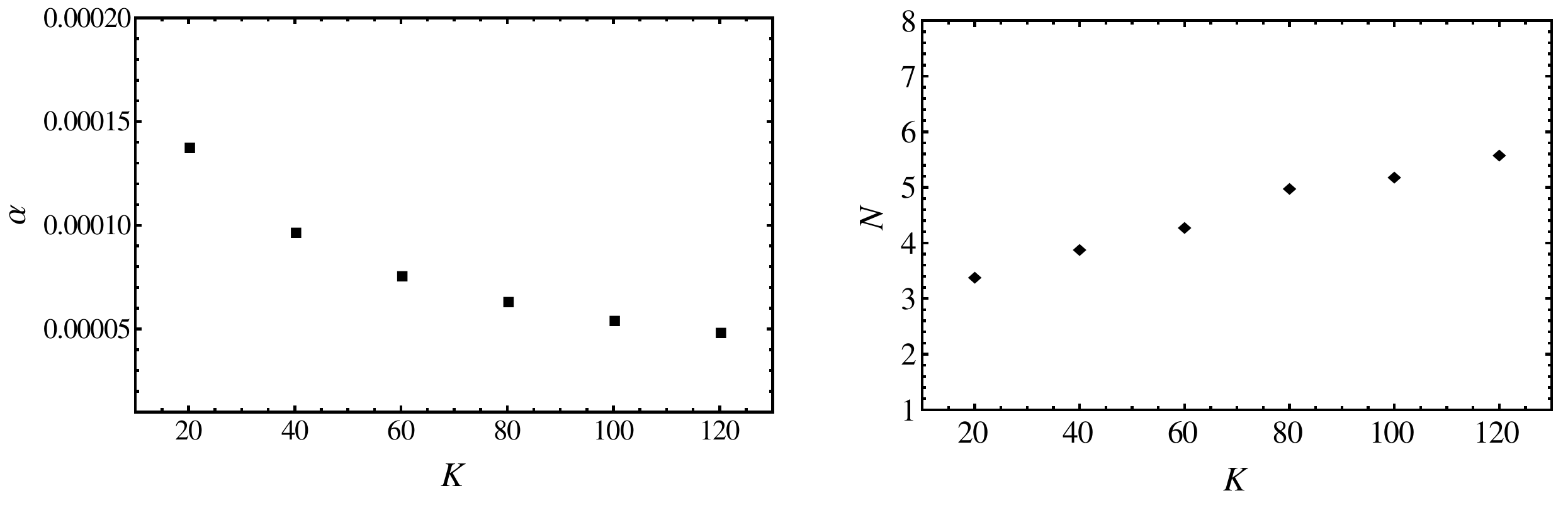}  
  \caption{The parameters $\alpha$ (left) and $N$ (right) in dependence of the portfolio size $K$ for daily returns, $\Delta t=1$ day, and optimal weights.}
\label{fig9}
\end{figure}

Till now, we have used 600 random portfolios of size $ K=20 $. What if we vary the number of portfolios or/and the number of stocks?
Increasing the number of portfolios reveals more of the tails of the histogram. Fig.~\ref{fig9} shows the case of varying the number of stocks $ K $. 
As the number of stocks $K$ increases, the value of $N$ increases too. The variance on the other hand decreases. In other words, we still see the benefit of diversification.

\section{Conclusion} \label{section5}

Non-stationarity is a common feature of financial data. We introduced an approach to model non-stationary correlations by an ensemble of Wishart correlation matrices. 
Averaging the multivariate normal distribution over this ensemble yields an elliptical distribution with exponential tails, which can be expressed in terms of a modified Bessel function of the second kind.

This approach allowed us to derive a pdf for the returns of a portfolio consisting of correlated stocks. For this purpose we assumed normally distributed asset returns at each fixed time step. This assumption is justified on short time scales, where the correlation matrix can be viewed as fixed.
The average portfolio return pdf depends only on two parameters: the bilinear form of the portfolio weights with the average empirical covariance matrix, which represents a portfolio variance, and a free parameter which 
characterizes the fluctuation strength of correlations.

The average portfolio return pdf describes the empirical data well, in particular in the central part of the distribution. This behavior is independent of the choice of portfolio weights.
Still, there are some deviations in the tails. This can be traced back to the fact that although the choice of the Wishart distribution for the random correlation matrix ensemble is indeed a reasonable assumption, it obviously can not capture all empirical details. Nevertheless, our results have a remarkable agreement with the data up to the third decade. The empirical evidence of log-normally rather than $\chi^2$ distributed variances suggest the choice of a log-normal distribution instead of (4).
This will be the subject of future work.

It is important to note that the average portfolio return distribution yields a good description of heavy-tailed portfolio return distributions with only one free parameter. It provides a better fit to the empirical data than the Student's $t$-distribution, 
which is one of standard heavy-tailed distributions in financial economics \citep[see e.g.,][]{Blattberg1974, Peiro1994}. 

To the best of our knowledge the presented ensemble approach is new in the economics literature. Recently, we also applied it in the context of credit risk \citep{Schmitt2014} leading to the computation of an averaged loss distribution for credit portfolios. 

Furthermore, the random matrix approach reduces the effect of fluctuating correlations to a single parameter characterizing their fluctuation strength. Thus, it enables us to estimate the fluctuation strength of correlations in a given time interval directly from the 
empirical return distributions without having to estimate the correlations on shorter time intervals. This is an important issue for portfolio optimization. 


\appendix

\section{} \label{appA}

We show that the pdf~(\ref{eq:multidist}) is well defined for $N < K $.  We diagonalize the covariance matrix in the following way 
\begin{align}
\Sigma=   U^\dagger \Lambda \ U
\end{align}
with an orthogonal matrix $ { U}={ U}^\dagger $ and a diagonal matrix of the eigenvalues $\Lambda$. In the case $ N<K $ the covariance matrix has $ N $ positive non-zero and $ K-N $ zero eigenvalues, $\Lambda=\diag (\lambda_1,\dots ,\lambda_N, 0,\dots, 0)$.
We rotate the vector $  x $ into the eigenbasis of the covariance matrix, write ${ v}={ U} { x}$, and obtain the pdf
\begin{align}
g (  v| \Lambda) &= \frac{ 1 }{ \sqrt{2 \pi }^K } \frac{ 1 }{ \sqrt{\det \Lambda } } \exp \left( - \frac{ 1 }{ 2 }  v^\dagger \Lambda^{-1}  v \right) \notag \\ 
&= \prod_{k=1}^{N} \frac{1}{\sqrt{2 \pi \lambda_k}}\exp \left(-\frac{v_k^2}{2 \lambda_k} \right) \prod_{k=N+1}^K \delta(v_k) \  ,
\end{align}
which clearly is well defined. A corresponding line of reasoning can be used for the Fourier transform~(\ref{eq:fouriertr}).

\section{} \label{appB}

We argue that 
\begin{align}
\widehat w ({A}| \Sigma,N)= \sqrt{ \frac{ N }{ 2 \pi } }^{ K N } \frac{ 1 }{ \sqrt{ \det \Sigma }^N } \ \exp \left(-\frac{ N }{ 2 } \trace  {A}^\dagger \Sigma^{-1} { A} \right)
\end{align}
is equivalent to $w ({ W}| { C},N)$, where $\Sigma= \sigma C\sigma$ and $ A=\sigma W $.
To this end we have to show that
\begin{align}
\int {d}[ { A} ] \ \widehat w ({ A}| \Sigma,N) \ f({ A})= \int { d}[ { W} ] \ w ({ W}| { C},N) \ f({ \sigma W})
\end{align}
with an arbitrary test function $f({ A})$. Changing the variables according to $ A=\sigma W $, we find
\begin{align}
\int { d}[ { A} ] \ \widehat w ({ A}| \Sigma,N) \  f({ A}) &= (\det \sigma)^N \int { d}[ { W} ] \ \widehat  w ({ \sigma W}| { \Sigma,N}) \ f({ \sigma W}) \\
&=\int { d}[ { W} ] \ w ({ W}| { C},N) \ f({ \sigma W}) \ ,
\end{align}
which proves the assertion.

\section{}  \label{appC}

We show the appearance of the $\chi^2$ representation in Eq.~(\ref{eq:result}), as it is reminiscent of stochastic volatility models. Using the formula 
\begin{align}
\frac{ 1 }{ a^\eta } = \frac{ 1 }{ 2^{\eta} \Gamma(\eta)} \int\limits^\infty_0 z^{\eta - 1} \exp \left( - \frac{a}{2} z \right) \, { d} z 
\end{align}
we can rewrite Eq.~(\ref{eq:result}) as 
\begin{align}
 \langle g \rangle ( x | \Sigma, N)= \int\limits_0^\infty \chi_N^2 (z) \ g \left(  x \bigg \vert \frac{z}{N} \Sigma \right) { d} z 
\end{align}
with the $\chi^2$ distribution of $N$ degrees of freedom
\begin{align}
 \chi_N^2 (z)=\frac{1}{2^{N/2} \Gamma (N/2)} z^{N/2-1} \exp \left( - \frac{z}{2} \right)
\end{align}
for $z\geq 0$ and zero otherwise.

\end{document}